\documentclass[a4paper, amsfonts, amssymb, amsmath, reprint, showkeys, nofootinbib, twoside,superscriptaddress,floatfix,aps, twocolumn]{revtex4}
\pdfoutput=1 
\usepackage[utf8]{inputenc}
\usepackage{graphicx} 
\usepackage{braket}
\usepackage[separate-uncertainty=true]{siunitx}
\usepackage{amsmath}
\usepackage{xcolor}

\begin{document}
\title{Realization of a $CQ_3$ Qubit: energy spectroscopy and coherence}

\author{B. Kratochwil}
\email{benekrat@phys.ethz.ch}
\affiliation{Department of Physics, ETH Zürich,  CH-8093 Zürich, Switzerland}
\author{J.V. Koski}
\affiliation{Department of Physics, ETH Zürich,  CH-8093 Zürich, Switzerland}
\author{A.J. Landig}
\affiliation{Department of Physics, ETH Zürich,  CH-8093 Zürich, Switzerland}
\author{P. Scarlino}
\affiliation{Department of Physics, ETH Zürich,  CH-8093 Zürich, Switzerland}
\author{J.C. Abadillo-Uriel}
\affiliation{Department of Physics, University of Wisonsin-Madison, Madison WI 53706, United States}
\author{C. Reichl}
\affiliation{Department of Physics, ETH Zürich,  CH-8093 Zürich, Switzerland}
\author{S. N. Coppersmith}
\affiliation{University of New  South Wales, Sydney, New South Wales, 2053, Australia}
\author{W. Wegscheider}
\affiliation{Department of Physics, ETH Zürich,  CH-8093 Zürich, Switzerland}
\author{Mark Friesen}
\affiliation{Department of Physics, University of Wisonsin-Madison, Madison WI 53706, United States}
\author{A. Wallraff}
\affiliation{Department of Physics, ETH Zürich,  CH-8093 Zürich, Switzerland}
\author{T. Ihn}
\affiliation{Department of Physics, ETH Zürich, CH-8093 Zürich, Switzerland}
\author{K. Ensslin}
\affiliation{Department of Physics, ETH Zürich,  CH-8093 Zürich, Switzerland}

\begin{abstract} 

The energy landscape of a single electron in a triple quantum dot can be tuned such that the energy separation between ground and excited states becomes a flat function of the relevant gate voltages. These so-called sweet spots are beneficial for charge coherence, since the decoherence effects caused by small fluctuations of gate voltages or surrounding charge fluctuators are minimized. We propose a new operation point for a triple quantum dot charge qubit, a so-called $CQ_3$-qubit, having a third order sweet spot. We show strong coupling of the qubit to single photons in a frequency tunable high-impedance SQUID-array resonator. In the dispersive regime we investigate the qubit linewidth in the vicinity of the proposed operating point. In contrast to the expectation for a higher order sweet spot, we there find a local maximum of the linewidth. We find that this is due to a non-negligible contribution of noise on the quadrupolar detuning axis not being in a sweet spot at the proposed operating point. While the original motivation to realize a low-decoherence charge qubit was not fulfilled, our analysis provides insights into charge decoherence mechanisms relevant also for other qubits.

\end{abstract}

\maketitle

\section{Introduction}
A single electron occupying two tunnel coupled quantum dots can be operated as a charge qubit \cite{vanWiel2002,Frey2012,Petersson2010,Gorman2005,Stocklauser2017,Bruhat2018}. Control parameters of this qubit are the inter-dot tunnel coupling $t$ as well as the detuning $\delta$ defined as the energy difference of the left and right dot electrochemical potentials. Noise protection to first-order in detuning is obtained by operating the qubit at $\delta = 0 $ \cite{Scarlino2019PRL}. This operation point is called a first-order "sweet spot" since the first derivative of the qubit energy with respect to the detuning parameter vanishes. At this point dephasing due to detuning noise is minimal \cite{Thorgrimsson2017} and charge qubit linewidths below \SI{3}{\mega\hertz} have been reported \cite{Scarlino2019,Scarlino2019PRL}.

Additional qubit control parameters are obtained by increasing the number of quantum dots in the linear array by one, i.e., by using a linear triple quantum dot (TQD) \cite{Landig2018,Russ2018,Busl2013, Gaudreau2006}. For this system the qubit parameters are the tunnel coupling $t_{\mathrm{L}}$ between the left and the middle and $t_{\mathrm{R}}$ between the right and the middle quantum dots, as well as the left to right dot asymmetry $\delta = \varepsilon_{\mathrm{L}} -\varepsilon_{\mathrm{R}}$ and the middle to outer dot detuning $E_{\mathrm{M}} = \varepsilon_{M} - \left(\varepsilon_{\mathrm{L}}+\varepsilon_{\mathrm{R}}\right)/2$. Here, $\varepsilon_{\mathrm{L}}$, $\varepsilon_{\mathrm{M}}$, and $\varepsilon_{\mathrm{R}}$ are the single-particle energies of electrons in the left, middle and right quantum dot, respectively.

One promising TQD qubit in terms of noise protection is the charge quadrupole qubit \cite{Friesen2017, Koski2019}. This single electron qubit utilizes the TQD ground and second excited state with the qubit excitation energy $E_{02}$, whereas the first excited state is a leakage state not connected to the other states by a quadrupole moment. The quadrupole qubit has recently been investigated experimentally  \cite{Koski2019} by strongly coupling it to a single photon in a superconducting microwave resonator. It has a single sweet spot at $\delta=E_{\mathrm{M}}=0$ in both detuning parameters, since at this point $\partial E_{02}/ \partial \delta = \partial E_{02}/ \partial E_{\mathrm{M}} = 0$. 
Improved coherence was detected operating the qubit on the quadrupolar axis $E_{\mathrm{M}}$ with $\delta = 0$ compared to operating the qubit on the detuning axis  $\delta$ with $E_{\mathrm{M}} = 0$. 

In this work we experimentally explore a different TQD qubit that hosts a single electron, called $CQ_3$-qubit. The device layout is the same as for the quadrupolar qubit \cite{Koski2019}, but here, the qubit states are chosen to be the ground and first excited state of the TQD system. For symmetric tunnel coupling, the qubit excitation energy $E_{01}$ possesses a third order sweet spot with respect to the detuning $\delta$ at $\delta=0$ and the specific value $E_{\mathrm{M}}=E_{\mathrm{M}}^{\mathrm{Opt}}$, meaning that $\partial E_{01} /\partial \delta = \partial^2 E_{01} /\partial \delta^2 = \partial^3 E_{01} /\partial \delta^3 = 0$ at this point (see Sec. \ref{sec:Theory} for details). 

For operating the $CQ_3$-qubit, the resonator is coupled to the left quantum dot (see Fig.~\ref{fig:Fig1}(a)), leading to a dipolar coupling between the qubit states. This is in contrast to the quadrupole qubit, where the resonator is coupled to the middle quantum dot \cite{Koski2019} in order to avoid dipolar coupling.
The $CQ_3$-regime has the potential advantage that the two logical qubit states are the two lowest energy levels and there is no intermediate leakage state as for the quadrupolar qubit \cite{Koski2019}. However, the $CQ_3$-qubit has no sweet spot in $E_{\mathrm{M}}$. Sacrificing the sweet spot in $E_{\mathrm{M}}$  for a higher order sweet spot in $\delta$ is useful when the dominant noise originates from charge fluctuations at large distances from the qubit \cite{Friesen2017}. We find this to be crucial for understanding the properties of charge noise in semiconductor devices.

In this paper, we start by presenting the theory of the $CQ_3$-qubit.
We then show measurements of the qubit-resonator system in the dispersive and resonant limits, investigate the qubit linewidth as a function of detuning $\delta$. We also develop a noise model explaining our experimental findings.

\section{Theory}
\label{sec:Theory}
\begin{figure}
    \centering
    \includegraphics{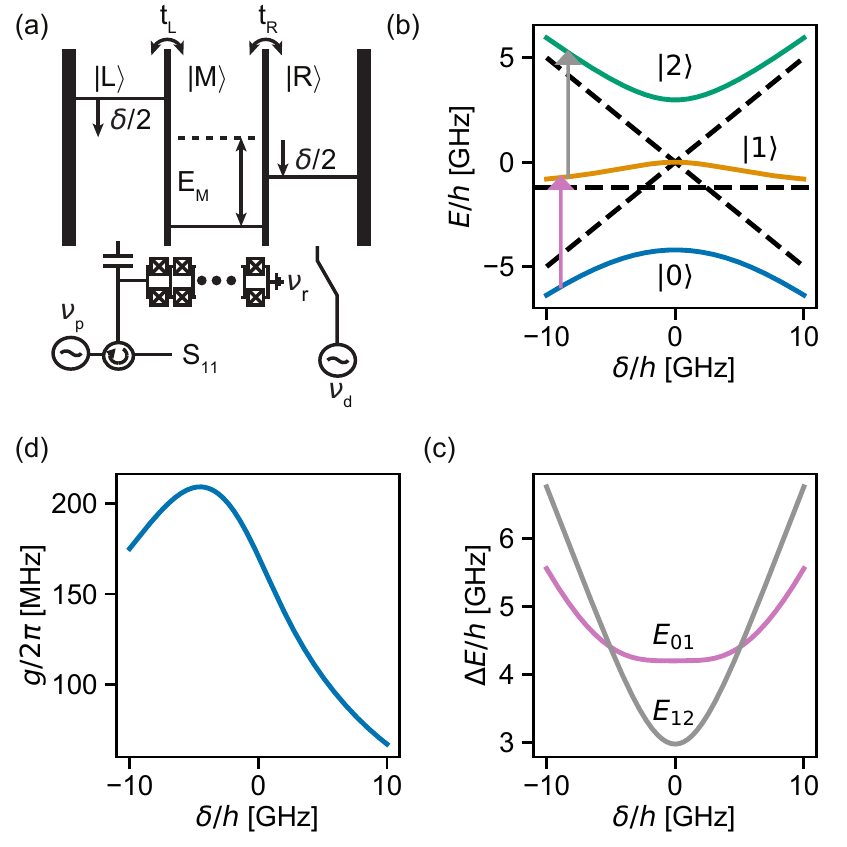}
    \caption{(a) Schematic diagram of the triple dot device. The left, middle and right quantum dots are indicated by $\ket{\mathrm{L}}$,$\ket{\mathrm{M}}$ and $\ket{\mathrm{R}}$ respectively. We can independently tune the chemical potentials and the inter-dot tunnel couplings $t_{\mathrm{L}}$ and $t_{\mathrm{R}}$ as well as the couplings to the reservoirs by electrostatic gating. A frequency tunable $\lambda/4$ SQUID array resonator is capacitively coupled to the left dot. We apply a drive tone to the qubit through the right dot plunger gate. (b) Spectrum of the Hamiltonian for equal tunnel couplings at $E_{\mathrm{M}} = E_{\mathrm{M}}^{\mathrm{Opt}}$ as a function of detuning $\delta$. Solid lines show the energy levels for $t/h = \SI{2.5}{\giga\hertz}$ and the dashed lines for  $t =0$. (c) Energy differences $E_{01}$ and $E_{12}$ of the spectrum from (b).  (d) Plot of the coupling strength of the resonator to the qubit as a function of detuning $\delta$.
    \label{fig:Fig1}}
\end{figure}

In the following we explore the Hamiltonian of a single electron confined in a TQD, as schematically shown in Fig.~\ref{fig:Fig1}(a). We first consider the bare qubit Hamiltonian neglecting coupling to the resonator. Subsequently we calculate the coupling matrix element to the resonator which is capacitively coupled to the left plunger gate.

In the position basis $\left\{\ket{\mathrm{L}}, \ket{\mathrm{M}}, \ket{\mathrm{R}}\right\}$, referring to an electron residing in either the left, middle or right dot, the Hamiltonian reads \cite{Friesen2017} 

\begin{equation}
    H = \begin{pmatrix}
            \delta/2 & t_{\mathrm{L}} & 0 \\
            t_{\mathrm{L}}^\star & E_{\mathrm{M}} & t_{\mathrm{R}} \\ 
            0 & t_{\mathrm{R}}^\star & -\delta/2
        \end{pmatrix}, \label{eq:H}
\end{equation}

where $t_{\mathrm{L}}$ and $t_{\mathrm{R}}$ describe the tunnel coupling between the middle-left and middle-right quantum dots, respectively. In the following we are interested in the symmetric coupling case, where $|t_{\mathrm{L}}| = |t_{\mathrm{R}}| = |t|$. Assuming that the quantity $|\delta/t|$ is small, we separate the Hamiltonian $H$ into the part $H_0=H(\delta=0,E_{\mathrm{M}})$, which can be diagonalized analytically, and the perturbation $H_1(\delta)=H-H_0$. We then perform second order perturbation theory to obtain the following approximate expression for the qubit excitation energy at $\mathcal{O}\left[\left|\delta /t \right|^2\right]$:
\begin{widetext}
    \begin{equation}
       E_{01} = -\frac{1}{2}\left(E_{\mathrm{M}} - \sqrt{E_{\mathrm{M}}^2 + 8|t|^2}\right) - \frac{E_{\mathrm{M}}^2+4|t|^2+3E_{\mathrm{M}} \sqrt{E_{\mathrm{M}}^2 + 8|t|^2}}{8|t|^2\sqrt{E_{\mathrm{M}}^2 + 8|t|^2}}\delta^2. \label{eq:E01}
    \end{equation}
\end{widetext}
In order to have a second order sweet spot $\left(\partial^2 E_{01}/\partial \delta^2|_{\delta = 0} = 0 \right)$ we set the prefactor of $\delta^2$ in \eqref{eq:E01} to zero. This leads to 
\begin{equation}
    E_{\mathrm{M}}^{\mathrm{Opt}} = -|t| \sqrt{3\sqrt{2} - 4} \approx -0.493\,|t|,
    \label{eq:emopt}
\end{equation}
where the point $(\delta=0, E_{\mathrm{M}}=E_{\mathrm{M}}^{\mathrm{Opt}})$ defines the $CQ_3$ operation point in the parameter space. By symmetry the third derivative $\partial^3 E_{01}/ \partial\delta^3$ also vanishes, yielding a third-order sweet spot in $\delta$ at this point. The energy levels of the qubit are plotted in Fig.~\ref{fig:Fig1}(b) as a function of $\delta$ at $E_{\mathrm{M}}=E_{\mathrm{M}}^{\mathrm{Opt}}$. The energy difference between the $\ket{0}$ and $\ket{1}$ states as well as the $\ket{1}$ and $\ket{2}$ states are plotted in Fig.~\ref{fig:Fig1}(c). We see a flat dispersion for $E_{01}$ around $\delta=0$, consistent with the third order sweet spot.

We next consider the qubit coupling to a resonator capacitively connected to the leftmost quantum dot, as indicated in Fig.~\ref{fig:Fig1}(a). We describe this coupling in the basis $\left\{ \ket{\mathrm{L}},\,\ket{\mathrm{M}}\, , \ket{\mathrm{R}} \right\}$ with the coupling Hamiltonian $H_{\mathrm{int}}=G(a^\dagger+a)$, where $a^\dagger$ and $a$ are the creation and annihilation operators for a photon in the cavity, and $G$ is the coupling matrix
\[ G = \hbar\omega_{\mathrm{r}}\sqrt{\frac{\pi Z}{h/e^2}}\left(\begin{array}{ccc}\alpha_{\mathrm{LM}} & 0 & 0 \\ 0 & 0 & 0 \\ 0 & 0 & \alpha_{\mathrm{RM}}\end{array}\right). \]
Here, $\hbar\omega_{\mathrm{r}}$ is the energy of a photon in the resonator, $Z$ is the resonator impedance, and $\alpha_{\mathrm{LM}}=\alpha^{\mathrm{L}}_{\mathrm{L}}-\alpha^{\mathrm{L}}_{\mathrm{M}}$ and $\alpha_{\mathrm{RM}}=\alpha^{\mathrm{L}}_{\mathrm{R}}-\alpha^{\mathrm{L}}_{\mathrm{M}}$ are the differences of the lever arms $\alpha^{\mathrm{L}}_j$ of the left plunger gate on dot $j$ ($j=\mathrm{L,M,R}$). The qubit resonator coupling is then found by transforming $G$ into its representation $\tilde{G}$ in the qubit basis $\ket{0}$, $\ket{1}$, $\ket{2}$. Given the matrix $S$ of eigenvectors of the Hamiltonian \eqref{eq:H}, which we compute numerically, the transformation is achieved by
\[ \tilde{G} = SGS^\dagger,\]
and the qubit--cavity coupling strength is
\[ g = \bra{0}\tilde{G}\ket{1}. \]
The resulting detuning-dependent coupling strength $g(\delta)$ is shown in Fig.~\ref{fig:Fig1}(d) for $E_{\mathrm{M}}=E_{\mathrm{M}}^{\mathrm{Opt}}$. For the plot we use $G_{\mathrm{LM}}/h \simeq \SI{203.4}{\mega\hertz}$ and  $G_{\mathrm{RM}}/h \simeq \SI{62.1}{\mega\hertz}$ as extracted from the experiment (see below).
Taking into account that the resonator couples more strongly to transitions between the left and the middle dot, we find that $g(\delta)$ exhibits a pronounced maximum at negative values of $\delta$.
It arises at the point where the electrochemical potentials of the left and middle dot are aligned, giving rise to a large dipole moment.

\section{Experimental Setup}

Figure~\ref{fig:Fig1}(a) shows the schematic of the sample. An optical micrograph can be found in the Appendix. The TQD is defined on a GaAs/AlGaAs heterostructure hosting a two-dimensional electron gas \SI{90}{\nano\metre} below the surface by applying voltages between nano-fabricated aluminum gates on the surface and the electron gas. We electrostatically control the tunnel couplings, as well as the electrochemical potentials of the dots, applying negative voltages to the corresponding gate electrodes. We measure the charge state of the TQD with a nearby quantum point contact (QPC). We can apply a drive tone at frequency $\nu_{\mathrm{d}}$ to the right plunger gate. The left dot plunger gate is capacitively coupled to a $\lambda/4$ SQUID-array resonator \cite{Stocklauser2017,Landig2019,Scarlino2019,vanWoerkom2018,Koski2019}. Changing the flux $\Phi$ threading the SQUID loops of the resonator, we can tune the resonator's bare resonance frequency by several \si{\giga\hertz}. The resonator impedance is $Z \simeq \SI{1.1}{\kilo\ohm}$. This enhances the resonator--qubit coupling strength by a factor of approximately $\SI{5}{}$ compared to standard $\SI{50}{\ohm}$ resonators. For all the experiments presented in this work the average photon number in the resonator is less than one (see the Appendix for details).

\section{Results}

\begin{figure}
    \centering
    \includegraphics{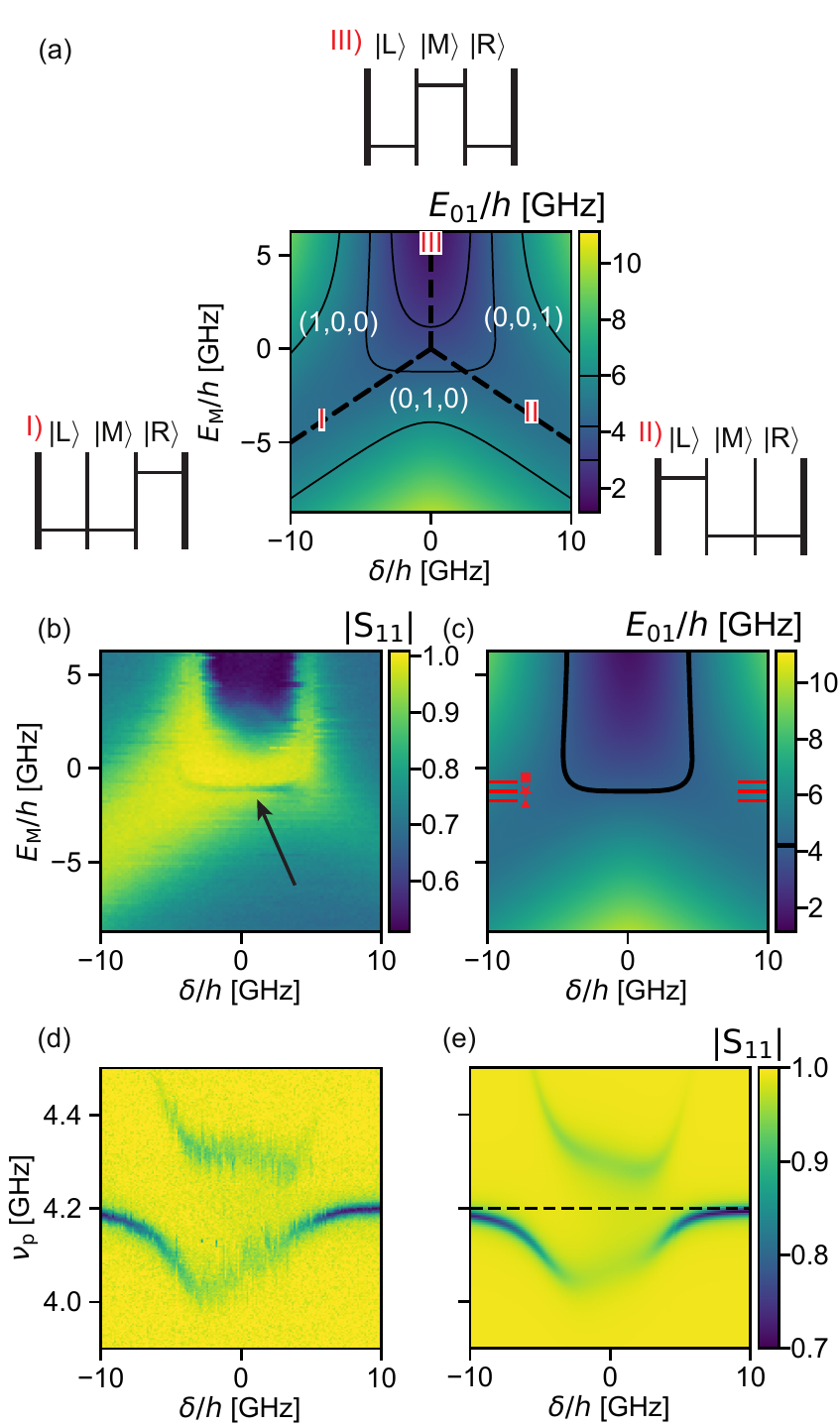}
    \caption{(a) TQD energy level schematic. Qubit energy $E_{01}/h$ for $|t|/h = \SI{2.5}{\giga\hertz}$ as a function of $\delta/h$ and $E_{\mathrm{M}}/h$. The thin solid black lines indicate qubit energy contours. The black dashed line show the charge transition lines, separating the single electron charge states of the qubit. Red roman numbers show the quantum dot energy level configurations at the indicated points in the configuration space. (b) Two tone spectroscopy measurement. Measured $|S_{11}|$ as a function of $\delta$ and $E_{\mathrm{M}}$ for $t/h = \SI{2.5}{\giga\hertz}$. A drive-tone is applied to the right plunger gate at $\nu_{\mathrm{d}} = \SI{4.2}{\giga\hertz}$ to map where $E_{01}/h = \nu_{\mathrm{d}}$. The resonator is at $\nu_{\mathrm{r}} = \SI{3.791}{\giga\hertz}$. (c) Calculated qubit energy as a function of $\delta$ and $E_{\mathrm{M}}$. The black line is the qubit energy contour for $E_{01}/h = \SI{4.2}{\giga\hertz}$. The three red lines and icons and symbols are referred to in Fig.~\ref{fig:Fig3}. (d) Resonant interaction of the qubit and the resonator for  $\nu_{\mathrm{r}}= \SI{4.2}{\giga\hertz}$ and $E_{\mathrm{M}} \approx  E_{\mathrm{M}}^{\mathrm{Opt}}\approx 0.493|t|$. (e) Simulation of $|S_{11}|$ for the measurement in (d) using Input-Output theory. The simulation parameters are $\nu_{\mathrm{r}} = \SI{4.19}{\giga\hertz}$, $\kappa_{\mathrm{int}}/2 \pi = \SI{14}{\mega\hertz} $, $\kappa_{\mathrm{ext}}/2 \pi= \SI{1.3}{\mega\hertz}$, $|t_{\mathrm{L}}|/h= \SI{2.47}{\giga\hertz} $ and $|t_{\mathrm{R}}|/h= \SI{2.48}{\giga\hertz} $. For the decoherence $\gamma_2$ we use the noise model we developed to fit the data in Fig.~\ref{fig:Fig4}(b). The black dashed line is a guide to the eye at \SI{4.2}{\giga\hertz}.
    \label{fig:Fig2}}
    
\end{figure}

In the following we experimentally investigate the qubit proposed above. We make use of the charge sensing QPC to tune the TQD into the single electron regime. The relevant charge states in the  $\left\{\ket{\mathrm{L}}, \ket{\mathrm{M}}, \ket{\mathrm{R}}\right\}$ basis are $\ket{1,0,0},\,\ket{0,1,0}$ and $\ket{0,0,1}$. We plot the qubit excitation energy as a function of the dipolar $\delta$ and quadrupolar $E_{\mathrm{M}}$ detuning in Fig.~\ref{fig:Fig2}(a) for $|t_{\mathrm{L}}|/h = |t_{\mathrm{R}}| /h= \SI{2.5}{\giga\hertz}$. The thin solid lines indicate contours of constant qubit excitation energy. The dashed black lines indicate the charge transition lines. We schematically depict the quantum dot energy levels at three points along the charge transition lines as indicated with roman numbers. The tunnel couplings $t_{\mathrm{L}}$ ($t_{\mathrm{R}}$) are determined by two tone spectroscopy of the DQD charge qubit formed between the left-middle (right-middle) dots at $\left|\delta\right|\gg t$ and negative $E_{\mathrm{M}}$, indicated by panels I and II in  Fig.~\ref{fig:Fig2}(a). These measurements also allow us to relate the qubit detuning parameters $\delta$ and $E_{\mathrm{M}}$ to combinations of plunger gate voltages by determining the gate lever arms (see Appendix). For the measurements presented in the following we set $|t_{\mathrm{L}}|/ h = |t_{\mathrm{R}}| /h= \SI{2.5}{\giga\hertz}$. The corresponding qubit energy $E_{01}$ at  $\delta=0$ and $E_{\mathrm{M}}=E_{\mathrm{M}}^{\mathrm{Opt}}$ is  $\SI{4.2}{\giga\hertz}$.

Next, we map out contour lines of the qubit energy using two tone spectroscopy \cite{Schuster2005}. We apply a drive tone $\nu_{\mathrm{d}} $ at $\SI{4.2}{\giga\hertz}$ to the right plunger gate while measuring the reflection $\left|S_{11}\right|$ of the probe tone applied at the bare resonator frequency $\nu_{\mathrm{p}} = \nu_{\mathrm{r}} = \SI{3.791}{\giga\hertz}$. The resonator and qubit interact off resonance, which leads to a dispersive shift of magnitude $\approx \pm g^2/\left(h \nu_{\mathrm{r}} - h\nu_{\mathrm{q}}\right)$ \cite{Schuster2005}.
The measured reflection as a function of the qubit detuning parameters $\delta$ and $E_{\mathrm{M}}$ is shown in Fig.~\ref{fig:Fig2}(b). On resonance the qubit excited state population increases, which leads to a decrease in the magnitude of the dispersive shift \cite{Schuster2005}, and consequently to a measurable change in  $\left|S_{11}\right|$. As expected for the higher order sweet spot discussed above, we find a flat dispersion along $\delta$,  which is indicated by an arrow.

We compare the result to theory by plotting the calculated qubit energy $E_{01}/h$ for  $|t_{\mathrm{L}}|/h = |t_{\mathrm{R}}|/h = \SI{2.5}{\giga\hertz}$ as a function of the qubit detuning parameters $\delta$ and $E_{\mathrm{M}}$ in Fig.~\ref{fig:Fig2}(c). The solid black line corresponds to the energy contour of \SI{4.2}{\giga\hertz}. It shows excellent agreement with the measurement in Fig.~\ref{fig:Fig2}(b).


Next we operate the qubit at the optimal working point $E_{\mathrm{M}} = E_{\mathrm{M}}^{\mathrm{Opt}} \approx -0.493 |t|$ (see Eq.~\ref{eq:emopt}) and probe the qubit on resonance $\nu_{\mathrm{r}} = E_{01}/h = \SI{4.2}{\giga\hertz}$ with the resonator. At constant resonator frequency $\nu_{\mathrm{r}}$ we change the qubit frequency by sweeping $\delta$ and measure the amplitude of the reflected resonator signal $\nu_{\mathrm{p}}$, as shown in Fig.~\ref{fig:Fig2}(d). As a consequence of the coherent qubit-photon hybridization we observe two resonance peaks in $|S_{11}|$ over a broad range of $\delta$. On resonance these two hybridized states with equal photon-matter character have an energy splitting of $2g$. The magnitude of this energy splitting changes as a function of $\delta$ in agreement with our theoretical model shown in Fig.~\ref{fig:Fig1}(d). The energy of the two states is approximately equally separated from the bare resonance frequency, indicating that the qubit energy is almost equal to that of the resonator (and therefore constant) for the whole range. This is an indication that the qubit dispersion is flat for a certain range in $\delta$ as one can see in Fig.~\ref{fig:Fig1}(c).

We simulate the reflectance spectrum $|S_{11}|$ of the qubit using an Input-Output model taking into account all relevant energy levels of the system \cite{Collett1985,Gardiner1984,Benito2017,Burkard2016} in Fig.~\ref{fig:Fig2}(e). We account for the detuning dependent coupling strength $g(\delta)$ as well as the detuning dependent decoherence rate discussed below (see Fig.~\ref{fig:Fig4}(c), details are found in the Appendix). It is important to note that a simple Jaynes Cummings model considering only the $\ket{0}$ and the $\ket{1}$ states would not reproduce the observed energy splitting of the  two resonances with the same parameters.

\begin{figure}
    \centering
    \includegraphics{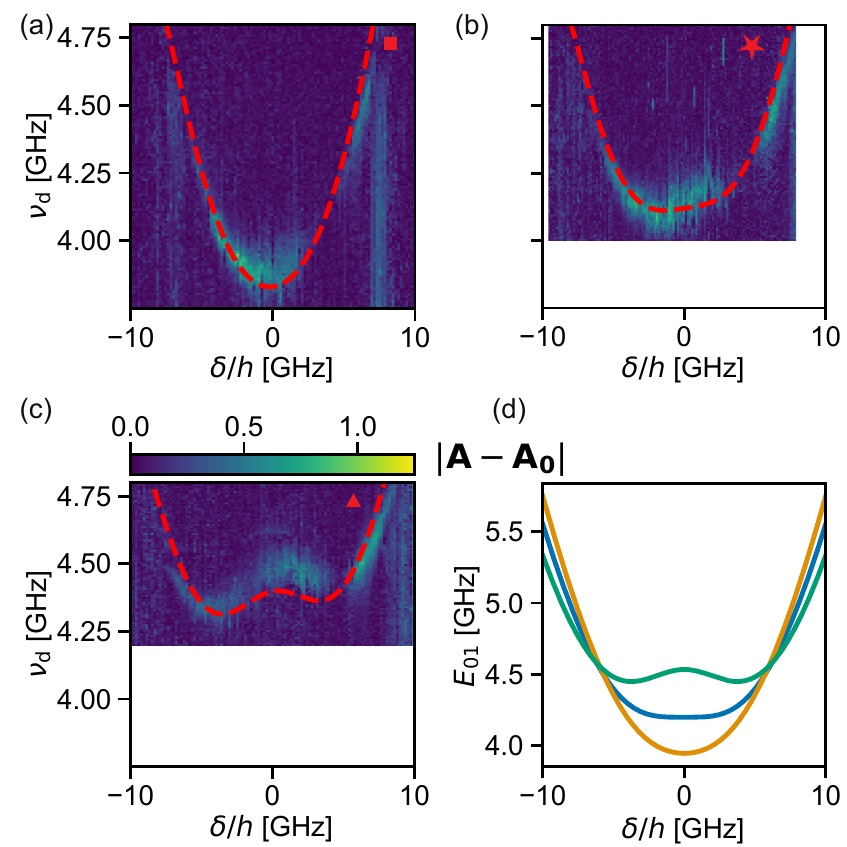}
    \caption{(a)-(c)Two tone spectroscopy of the qubit dispersion as a function of detuning $\delta$. We plot the complex amplitude $\left|A-A_{0}\right|$ of the reflected signal as a function of the drive frequency $\nu_{\mathrm{d}}$ and the detuning $\delta$. The red dashed lines show the expected qubit dispersion. We take the dispersive shift and the detuning dependent coupling $g(\delta)$ into account. The icons in the upper right corner correspond to the values for $E_{\mathrm{M}}$ at which the linecuts were made in Fig.~\ref{fig:Fig2}(b). (d) Plot of the qubit energy for three settings $E_{\mathrm{M}} = \left\{E_{\mathrm{M}}^{\mathrm{Opt}}/h+\SI{0.5}{\giga\hertz},E_{\mathrm{M}}^{\mathrm{Opt}}, E_{\mathrm{M}}^{\mathrm{Opt}}/h-\SI{0.5}{\giga\hertz}\right\}$ of the middle dot potential.
    \label{fig:Fig3}}
\end{figure}

We now perform two tone spectroscopy in the dispersive limit. For this purpose we tune the resonator frequency to $\nu_{\mathrm{r}} = \SI{5.1}{\giga\hertz}$ and keep the tunnel couplings at $|t|/h = \SI{2.5}{\giga\hertz}$. Sweeping the drive tone frequency $\nu_{\mathrm{d}}$ applied to the right plunger gate, and stepping $\delta$, we measure the complex amplitude $A$ of the reflected probe tone at frequency $\nu_{\mathrm{p}}$. We show the qubit dispersion as a function of detuning $\delta$ for three values of $E_{\mathrm{M}}$ in Figs.~\ref{fig:Fig3}(a-c). The measured dispersions correspond to three horizontal linecuts indicated by red lines in Fig.~\ref{fig:Fig2}(c) at $E_{\mathrm{M}} = \left\{E_{\mathrm{M}}^{\mathrm{Opt}}/h+\SI{0.5}{\giga\hertz},E_{\mathrm{M}}^{\mathrm{Opt}}, E_{\mathrm{M}}^{\mathrm{Opt}}/h-\SI{0.5}{\giga\hertz}\right\}$. The red dashed lines show the expected qubit dispersion according to calculation of $E_{01}$ also including the detuning dependent dispersive shift in the calculations. We see a good agreement with theory. A plot of the expected qubit dispersion not taking the dispersive shift into account is shown in Fig.~\ref{fig:Fig3}(d).

For $E_{\mathrm{M}}> E_{\mathrm{M}}^{\mathrm{Opt}}$ (Fig.~\ref{fig:Fig3}(a)) the dispersion has a single minimum, originating from a DQD charge qubit formed between the left and right quantum dot. The middle quantum dot acting as a tunnel barrier. This is schematically depicted in Fig.~\ref{fig:Fig2}(a), energy diagram III. The measured dispersion at $E_{\mathrm{M}}^{\mathrm{Opt}}$ (Fig.~\ref{fig:Fig3}(b)) is not completely flat in detuning, but slightly tilted due to the changing coupling strength as a function of detuning. We observe a vanishing visibility of the spectroscopic signal strength around the points where $E_{01} = E_{12}$. Furthermore, the linewidth of the signal strongly increases around $\delta = 0$. This is in contrast to the intuition of a higher order sweet-spot protected from decoherence and thus a narrower qubit linewidth \cite{Thorgrimsson2017}. For $E_{\mathrm{M}} < E_{\mathrm{M}}^{\mathrm{Opt}}$ (Fig.~\ref{fig:Fig3}(c)) the dispersion has two distinct local minima that arise from the DQD forming between the left-middle (right-middle) DQD for negative (positive) $\delta$ respectively. Furthermore, we observe a flickering in the signal when approaching the predicted sweet spot. We observed this behavior in resonant interaction (see Fig.~\ref{fig:Fig2}(d)) as well as in two tone spectroscopy measurements (see Fig.~\ref{fig:Fig3}(a-c)). The same behavior was also found when investigating the qubit at other tunnel coupling configurations. A possible explanation could be a fluctuating charge defect in the GaAs/AlGaAs heterostructure capacitively coupled to the TQD.


To further investigate the broadening of the qubit linewidth around $\delta = 0$  we measure the qubit linewidth at the optimal working point as a function of applied drive power, see Fig.~\ref{fig:Fig4}(a). The qubit frequency is $\nu_{\mathrm{q}} = \SI{4.2}{\giga\hertz}$ and the resonator frequency is $\nu_{\mathrm{r}} = \SI{5.1}{\giga\hertz}$. We measure the reflected signal $\nu_{\mathrm{p}}$  while stepping the drive tone $\nu_{\mathrm{d}}$ through resonance with the qubit at different drive tone powers $P_{\mathrm{d}}$. The half width at half maximum $\partial\nu_{\mathrm{q}}$ of the resonance depends on the applied drive tone power $P_{\mathrm{d}}$ according to $\partial \nu_{\mathrm{q}} = \sqrt{\left(\gamma_2/2\pi\right)^2 + \beta P_{\mathrm{d}}}$ \cite{Abragam1961,Schuster2005}, where $\beta$ is a constant describing the total attenuation along the drive line. The term $\beta P_{\mathrm{d}}$ describes the power broadening of the qubit linewidth. We plot the extracted $\partial\nu_{\mathrm{q}}^2$ as a function of drive power and observe the typical linear evolution of the linewidth as a function of applied power \cite{Stocklauser2017,Schuster2005,Landig2018,Scarlino2019}. We attribute the deviations from the linear evolution of the linewidth to the flickering observed in the previous measurements (Fig.~\ref{fig:Fig2}(d)Fig.~\ref{fig:Fig3}(a)-(c)). From the linear extrapolation of the data points to zero drive power we extract the decoherence rate $\gamma_2/2\pi = \SI{53 \pm 2}{\mega\hertz}$ for the qubit. 


\begin{figure}
    \centering
    \includegraphics{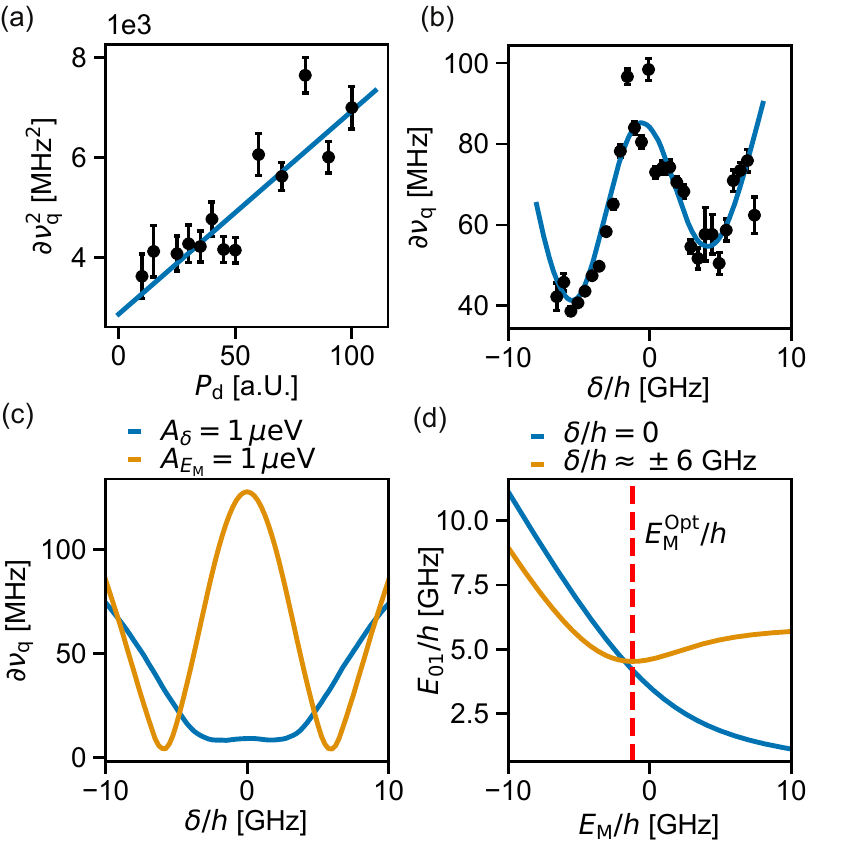}
    \caption{(a) Measurement of the half width half maximum squared vs applied probe power. The blue line is a fit to the measurement. We extract a qubit decoherence rate $\gamma_2 = \SI{53 \pm 2}{\mega\hertz}$.(b) Measurement of the half width half maximum of the qubit resonance as a function of detuning $\delta$. The blue line shows the fit using a noise model taking charge noise and magnetic noise into account. (c) Simulated qubit linewidth from 1/f charge noise of amplitude \SI{1}{\micro\eV} on either $\delta$ or $E_{\mathrm{M}}$. (d) Plot of $E_{01}/h$ as a function of $E_{M}$ for $\delta  = 0$ and $|\delta|/h \approx \SI{6}{\giga\hertz}$. The red dashed line indicates the qubit operation point.
    \label{fig:Fig4}}
\end{figure}

We further analyze the noise suffered by the qubit by measuring $\partial\nu_{\mathrm{q}}$ for a fixed finite drive power as a function of detuning $\delta$, see Fig.~\ref{fig:Fig4}(b)\cite{Scarlino2019PRL}. The drive tone is applied via the right plunger gate, the power broadening for positive detuning $\delta$ is therefore higher than for negative detuning. This causes a monotonic offset in the measured linewidth from negative to positive detuning caused by the detuning dependence of the wave functions entering the dipole moment. As already seen in Fig.~\ref{fig:Fig3}(a-c) we observe that at $\delta = 0 $ the qubit linewidth $\partial\nu_{\mathrm{q}}$ has a local maximum. We observe the minimal linewidth at $|\delta|/h \approx \SI{6}{\giga\hertz}$. Additionally, we find that the minimum for negative detuning $\delta$ is lower than for positive $\delta$.

The blue line in Fig.~\ref{fig:Fig4}(b) shows the results from a fit using a noise model taking 1/f charge-noise and magnetic noise into account \cite{Yang2019,Fei2015}. 
In order to understand the evolution of the qubit linewidth as a function of $\delta$ shown in Fig.~\ref{fig:Fig4}(b), we investigate the different contributions building up the noise spectrum of the qubit. The main noise source is charge noise acting either on $\delta$ or $E_{\mathrm{M}}$. Fig.~\ref{fig:Fig4}(c) shows the qubit linewidth simulated with a 1/f charge noise model, considering first  $\delta$ noise with spectral density  $S_\delta(\omega)=A_\delta^2/\omega$ and amplitude $A_{\delta} = \SI{1}{\micro\eV}$ (blue trace), or considering $E_{\mathrm{M}}$ noise with $S_{E_{\mathrm{M}}}(\omega)=A_{E_{\mathrm{M}}}^2/\omega$ and amplitude $A_{E_{\mathrm{M}}}= \SI{1}{\micro\eV}$ (brown trace). In case the system is affected by pure dipolar detuning noise the simulation shows a flat decoherence rate as a function of detuning, being in agreement with the higher order sweet spot. The finite linewidth at $\delta = 0$ can mostly be attributed to leakage to the second excited state. If, however, the system is affected only by noise in $E_{\mathrm{M}}$ there is a local maximum at $\delta = 0$, because the proposed qubit operation point has no sweet spot in $E_{\mathrm{M}}$. The two minima of the linewidth in Figs.~\ref{fig:Fig4}(b,c) close to $|\delta|/h \approx  \SI{6}{\giga\hertz}$ are due to sweet spots along the $E_{\mathrm{M}}$ axis.

A plot of the qubit energy $E_{01}$ as a function of $E_{M}$ is shown in Fig.~\ref{fig:Fig4}(d). The linecut at $\delta = 0$ has a non-vanishing slope whereas the linecut at $|\delta|/h \approx  \SI{6}{\giga\hertz}$ shows a local minimum at $E_{\mathrm{M}}^{\mathrm{Opt}}$ indicating the DQD qubit sweet spot. For further details we refer to the Appendix. We conclude that the maximum around $\delta = 0$ originates from a non-negligible noise contribution along the $E_{\mathrm{M}}$ axis. As mentioned above the main contribution to the asymmetry of the measured linewidth presented in Fig.~\ref{fig:Fig4}(b) is due to different power broadening for negative/positive detuning $\delta$.

For completeness we also consider the effect of magnetic noise due to different Overhauser fields in the dots \cite{Fei2015}. A plot for Gaussian-distributed magnetic noise either on the left-middle or the right-middle DQD is shown in Fig.~\ref{fig:FigS6} in the appendix. A model considering a combination of charge and magnetic noise, together with a constant detuning shift $\delta'$ (which can arise due to a change of the electrostatic environment of the qubit) shows good agreement with the measurement. From a fit we find: $A_{\delta} = \SI{1.949\pm 0.098}{\micro\eV}$, $A_{E_{\mathrm{M}}} = \SI{0.935\pm0.026}{\micro\eV}$, $\mathrm{corr} = \SI{-0.084\pm 0.177}{}$, $\delta' = \SI{0.69\pm0.171}{\micro\eV}$. The magnitude of $\delta'$ is plausible and within the range of electrostatic jumps observed during the measurements. The magnitude of the values found for  $A_{\delta}$, $A_{E_{\mathrm{M}}}$ and $\mathrm{corr}$ are in agreement with our previous work \cite{Koski2019}. Although we find that $A_{E_{\mathrm{M}}} < A_{\delta}$ in this experiment, we have shown that the former is still strong enough to dominate other decoherence mechanisms in the device. This is reasonable because the $CQ_3$ qubit was specifically designed to be protected from noise in the $\delta$ parameter only. These results support a growing body of evidence suggesting that a significant fraction of charge noise in semiconductor qubits originates from sources in the immediate vicinity of the quantum dot \cite{Koski2019,Connors2019}.

\section{Conclusion}

In conclusion we have proposed and measured a single-electron qubit hosted in a triple quantum dot with a third order sweet spot in the detuning parameter $\delta$. Using two-tone spectroscopy we mapped the qubit energy contour as a function of the two qubit parameters $\delta$ and $E_{\mathrm{M}}$. We observed a well resolvable vacuum Rabi splitting when bringing the cavity and the qubit into resonance. The reflected cavity signal was calculated using Input-Output theory taking all three qubit levels into account. With two tone spectroscopy we mapped out the qubit dispersion for different values of the middle dot potential $E_{\mathrm{M}}$ and found good agreement with calculations of the energy spectra. The qubit is expected to have reduced sensitivity to charge noise in the dot detuning. At the same time, the energy dispersion as a function of middle dot energy is not flat, making the system susceptible to quadrupolar charge noise. Experimentally we observe that the qubit linewidth around $\delta = 0 $ is maximal, proving that quadrupolar noise cannot be neglected in this system. We extract a decoherence rate $\gamma_2 = \SI{53 \pm 2}{\mega\hertz}$ in the limit of zero applied drive power. We further investigate the qubit linewidth as a function of $\delta$. We observe a maximum in linewidth at $\delta = 0 $ and two local minima at $|\delta|/h \approx  \SI{6}{\giga\hertz}$. We find good agreement to a 1/f charge noise model also considering magnetic noise due to hyperfine interaction. These results indicate that the noise affecting the qubit has a non-negligible contribution coming from short-range noise sources, in contrast to the original hypothesis that noise mostly originates from long-ranged sources.

\section*{Acknowledgments}

We acknowledge fruitful discussions with Christian Kraglund Andersen. This work was supported by the ETH FIRST laboratory and the Swiss National Science Foundation through the National Center of Competence in Research (NCCR) Quantum Science  and  Technology. SNC and MF acknowledge support by the Vannevar Bush Faculty Fellowship program sponsored by the Basic Research Office  of the  Assistant  Secretary of Defense for Research and Engineering and funded by the Office of Naval Research through Grant No. N00014-15-1-0029. MF and JCAU acknowledge support by ARO (W911NF-17-1-0274). The views and conclusions contained herein are those of the authors and should not be interpreted as necessarily representing the official policies or endorsements, either expressed or implied, of the Army Research Office (ARO) or the U.S. Government. The U.S. Government is authorized to reproduce and distribute reprints for Governmental purposes, notwithstanding any copyright notation thereon.
\section{Appendix}
\subsection{Sample}

\begin{figure}
    \centering
    \includegraphics[width=\columnwidth]{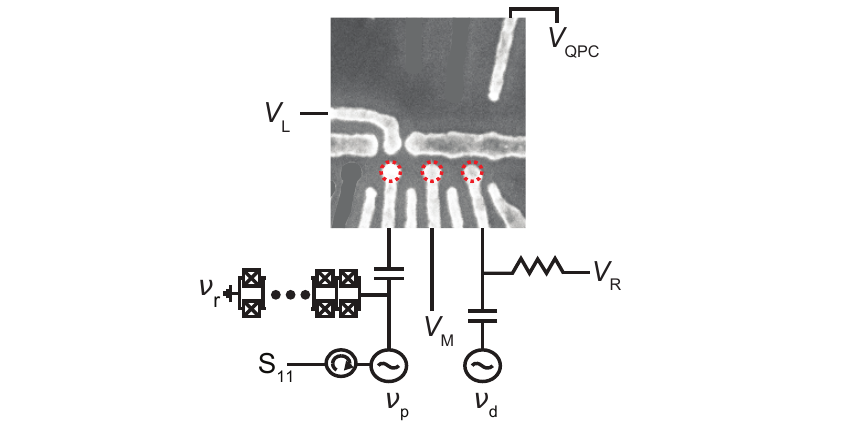}
    \caption{Scanning electron micrograph picture of the sample. The three dots are indicated by the red dashed circles. We control the electrochemical potentials of the dots by tuning the voltages $V_{\mathrm{L}}$, $V_{\mathrm{M}}$, $V_{\mathrm{R}}$ applied to the corresponding plunger gates. The inter dot tunnel couplings as well as the coupling to the reservoirs is controlled by the rest of the gates. The left plunger gate that directly overlaps the dot is capacitively connected to the $\lambda/4$ resonator keeping its potential grounded. To change the left dot potential we tune the potential $V_{\mathrm{L}}$. Changing $V_{\mathrm{QPC}}$ we tune the electrostatic potential of the charge sensing QPC. 
    \label{fig:FigS1}}
\end{figure}

We present a scanning electron image of the sample in Fig.~\ref{fig:FigS1}. Gates which are not used are grayed out. The three quantum dots reside under the plunger gates and are indicated by the red dashed circles. The plunger gate of the left quantum dot is capacitively coupled to the $\lambda/4$ resonator. Since the resonator DC potential is defined by the ground, the DC potential of the left quantum dot is tuned by the gate potential $V_{\mathrm{L}}$ as indicated in Fig.~\ref{fig:FigS1}.  Therefore its potential can not be tuned by applying a gate voltage. To tune the potential of the left dot we use the additional gate coming from the top. To check the occupation of the dots we use the nearby quantum point contact \cite{Schleser2004,Gaudreau2006}.

\subsection{Calibrating the qubit}

In the following we present how we tune the TQD into the correct regime and calibrate the relevant qubit control parameters. We start by tuning the TQD into the correct charge state. In the next step we introduce the voltages $V_{\delta}$ and $V_{\mathrm{E}_{\mathrm{M}}}$ which tune the potentials of the dots in a symmetric or anti-symmetric fashion, respectively. In the last step we present how we calibrate the tunnel couplings $t_{\mathrm{L(R)}}$, respectively, and how we convert the measurement axis into frequency space.

Using the QPC we tune the TQD into the single electron regime. In the basis $\left\{\ket{\mathrm{L}},\ket{\mathrm{M}},\ket{\mathrm{R}}\right\}$ of the left, middle and right dot, the relevant charge states in the occupation number representation are $\ket{1,0,0}$, $\ket{0,1,0}$ and $\ket{0,0,1}$, in other words having one electron in one of the three dots. In the first step we tune the qubit by directly tuning the plunger gate voltages $V_{\mathrm{x}}, \mathrm{with } \;\mathrm{x} \in \left\{\mathrm{L},\mathrm{M},\mathrm{R}\right\}$. In the next step we parametrize the plunger gate voltages by $V_{\delta}$ and $V_{E_{\mathrm{M}}}$, corresponding to symmetric and anti-symmetric voltage changes in the TQD. The change in $V_{\mathrm{M}}$ is determined by measuring two charge stability diagrams as function of $V_{\mathrm{L}}$ and $V_{\mathrm{R}}$ where we decrease $V_{M}$ by \SI{1}{\milli\volt}. We find that the quadruple point of the charge states  $\ket{0,1,0}$, $\ket{1,0,0}$, $\ket{0,0,1}$, $\ket{1,0,1}$ shifts by \SI{1.5}{\milli\volt} and \SI{0.4}{\milli\volt} in $V_{\mathrm{L}}$ and $V_{\mathrm{R}}$ respectively. The lever arms for $V_{\delta}$ are determined by measuring a charge stability diagram as function of $V_{\mathrm{L}}$ and $V_{\mathrm{R}}$ and calculating the tilt of the $\ket{0,1,0}$ and $\ket{1,0,1}$ transition when measured in the $V_{\delta}$ and $V_{E_{\mathrm{M}}}$ basis. From these measurements we find:

\begin{figure}
    \centering
    \includegraphics[width=\columnwidth]{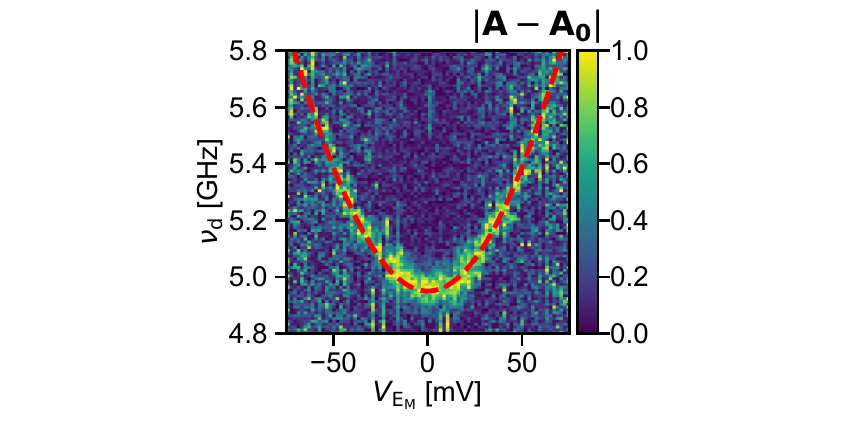}
    \caption{
    Two tone spectroscopy measurement of the middle-right DDQ charge qubit. We keep $V_{\delta}$ constant while measuring the qubit dispersion along $V_{\mathrm{E}_{\mathrm{M}}}$. The red line shows a fit to the measurement. From this we extract the tunnel coupling $t_{\mathrm{R}}$ and the lever arm $\alpha_{\mathrm{MR}}^{\mathrm{M}}$. 
    \label{fig:FigS3}}
\end{figure}

\begin{align}
    V_{\mathrm{L}} =& V_{\mathrm{L},0} + 1.5V_{E_{\mathrm{M}}} - V_{\delta}\\
    V_{\mathrm{M}} =& V_{\mathrm{M},0} - V_{E_{\mathrm{M}}}\\
    V_{\mathrm{R}} =& V_{\mathrm{R},0} +0.4 V_{E_{\mathrm{M}}} + 0.6 V_{\delta},
\end{align}

where $V_{\mathrm{x},0}, \mathrm{with} \,\mathrm{x} \in \left\{\text{L,M,R}\right\}$ are constants and chosen such that higher electron occupation numbers of the TQD are not relevant when operating the qubit. This parameterization is device-dependent and is determined by the inter dot and gate capacitances \cite{vanWiel2002}. It is interesting to note however, that the lever arms for $V_{\mathrm{L}}$ are lower than for $V_{\mathrm{R}}$ being in agreement with the fact that the plunger gate is further away from the dot, see Fig.~\ref{fig:FigS1}.

In a next step we want to calibrate the tunnel couplings $t_{\mathrm{L}}$ and $t_{\mathrm{R}}$ of the qubit and relate voltage changes in  $V_{E_{\mathrm{M}}}$ and $V_{\delta}$ to the physical qubit detuning parameters $\delta$ and $E_{\mathrm{M}}$. By tuning $V_{E_{\mathrm{M}}}$ more positive we can operate the device like a conventional DQD charge qubit formed between the left-middle (middle-right) DQD for large negative (positive) detuning voltage $V_{\delta}$, see Fig.~\ref{fig:Fig2}(a) energy diagram I and II. We map the DQD dispersion along $V_{\delta}$ and $V_{E_{\mathrm{M}}}$ for both the left-middle and right-middle  DQDs. We show the measurement of the left-middle DQD as a function of $V_{E_{\mathrm{M}}}$ in Fig.~\ref{fig:FigS3}. The red line shows a fit to the data using the DQD dispersion \cite{vanWiel2002}
\begin{equation}
    E_{\mathrm{DQD}} = \sqrt{\left(\alpha_{\mathrm{xy}}\left(V-V_{0}\right)\right)^2 + (2|t|)^2},
\end{equation}
where $|t|$ is the tunnel coupling of the respective DQD, $V_0$ is the voltage at which the dispersion has its minimum, $V$ is the corresponding detuning voltage and $\alpha_{\mathrm{xy}}$ is the lever arm that converts the voltage into a frequency. The energy at the minimum of the dispersion is given as  $2t$. Using the gate electrodes we tune $2|t| = \SI{5}{\giga\hertz}$. From the fit of the frequency response measurement discussed in the main text we find $|t_{\mathrm{L}}| = \SI{2.47}{\giga\hertz}$ and $|t_{\mathrm{R}}| = \SI{2.48}{\giga\hertz}$. This change of approximately \SI{2}{\percent} is attributed to the change of tunnel couplings when changing the detuning parameters. 

In the last step we want to calibrate the voltage changes $V_{\delta}$ and $V_{E_{\mathrm{M}}}$ to detuning changes in $\delta$ and $E_{\mathrm{M}}$. From the above mentioned two tone spectroscopy measurements we extract lever arms $\alpha_{\mathrm{xy}}$. More specifically from two tone spectroscopy on the left-middle DQD we extract $\alpha_{\mathrm{ML}}^{\delta}$ from spectroscopy along $V_{\delta}$ and $\alpha_{\mathrm{ML}}^{\mathrm{M}}$ from spectroscopy along $V_{E_{\mathrm{M}}}$. For further calculations we define the following detuning parameters which we later will relate to the measured lever arms. 

\begin{align}
    \delta_{\mathrm{LR}} =& \varepsilon_{\mathrm{L}} - \varepsilon_{\mathrm{R}} = \delta \\
    \delta_{\mathrm{MR}} =& \varepsilon_{\mathrm{M}} - \varepsilon_{\mathrm{R}}\\
    \delta_{\mathrm{ML}} =& \varepsilon_{\mathrm{M}} - \varepsilon_{\mathrm{L}}\\
    E_{\mathrm{ML}}^{\mathrm{M}} =& \varepsilon_{\mathrm{M}}- \varepsilon_{\mathrm{L}}\\
    E_{\mathrm{MR}}^{\mathrm{M}} =& \varepsilon_{\mathrm{M}}- \varepsilon_{\mathrm{R}}
\end{align}

Voltage changes $\Delta V_{\delta}$ and $\Delta V_{E_{\mathrm{M}}}$ relate to the measured lever arms according to

\begin{align}
    \Delta\delta_{\mathrm{XY}} =&  \alpha^{\delta}_{\mathrm{XY}} \Delta V_{\delta}\\
    \Delta   E_{\mathrm{XY}}^{\mathrm{M}} =& \alpha^{\mathrm{M}}_{\mathrm{XY}} \Delta V_{E_{\mathrm{M}}}.
\end{align}

We obtain the relevant lever arms for the TQD by forming the linear combinations
\begin{align}
    \alpha^{\delta}_{\mathrm{LR}} =& \alpha^{\delta}_{\mathrm{MR}}-\alpha^{\delta}_{\mathrm{ML}},\\
    \alpha^{\mathrm{M}}_{\mathrm{LR}} =& \alpha^{\mathrm{M}}_{\mathrm{MR}} - \alpha^{\mathrm{M}}_{\mathrm{ML}},\\
    \alpha^{\delta}_{E_{\mathrm{M}}} =& \frac{1}{2}\left(\alpha^{\delta}_{\mathrm{MR}}+\alpha^{\delta}_{\mathrm{ML}}\right),\\
    \alpha^{\mathrm{\mathrm{M}}}_{E_{\mathrm{M}}} =& \frac{1}{2}\left(\alpha^{\mathrm{M}}_{\mathrm{MR}}+\alpha^{\mathrm{M}}_{\mathrm{ML}}\right).\\
\end{align}    

Introducing normalization parameters $A$ we obtain the following group of linear equations: 

\begin{equation}
  \begin{pmatrix}
  \SI{1}{\giga\hertz}\\
  0\\
  0\\
   \SI{1}{\giga\hertz}
  \end{pmatrix} = 
  \begin{pmatrix}
  \alpha^{\delta}_{\mathrm{LR}} &  \alpha^{\mathrm{M}}_{\mathrm{LR}} & 0 & 0 \\
  \alpha^{\delta}_{E_{\mathrm{M}}} & \alpha^{\mathrm{M}}_{E_{\mathrm{M}}} & 0 & 0 \\
  0 & 0 & \alpha^{\delta}_{\mathrm{LR}} & \alpha^{\mathrm{M}}_{\mathrm{LR}}\\
  0 & 0 & \alpha^{\delta}_{E_{\mathrm{M}}} & \alpha^{\mathrm{M}}_{E_{\mathrm{M}}} \\
  \end{pmatrix} 
  \begin{pmatrix}
  A_{\mathrm{X}\delta}\Delta V_{\delta}\\ 
  A_{\mathrm{XM}}\Delta V_{E_{\mathrm{M}}} \\
  A_{\mathrm{Y}\delta}\Delta V_{\delta}\\
  A_{\mathrm{YM}}\Delta V_{E_{\mathrm{M}}}\\
  \end{pmatrix}
\end{equation}

Rewriting this equation in more compact form we find:

\begin{multline}
    \begin{pmatrix}
     \SI{1}{\giga\hertz} & 0 \\ 
    0 &  \SI{1}{\giga\hertz}\\
    \end{pmatrix}\\ = 
    \begin{pmatrix}
    \alpha^{\delta}_{\mathrm{LR}} &  \alpha^{\delta}_{\mathrm{LR}}\\
    \alpha^{\delta}_{E_{\mathrm{M}}} & \alpha^{\mathrm{M}}_{E_{\mathrm{M}}}\\
    \end{pmatrix}
    \begin{pmatrix}
    A_{\mathrm{X}\delta}\Delta V_{\delta} &  A_{\mathrm{Y}\delta}\Delta V_{\delta}\\ 
    A_{\mathrm{XM}}\Delta V_{E_{\mathrm{M}}} & A_{\mathrm{YM}}\Delta V_{E_{\mathrm{M}}} \\
    \end{pmatrix}
\end{multline}

By solving this equation we end up with the desired lever arms.

\subsection{Input-Output Theory}

\begin{figure}
    \centering
    \includegraphics[width=\columnwidth]{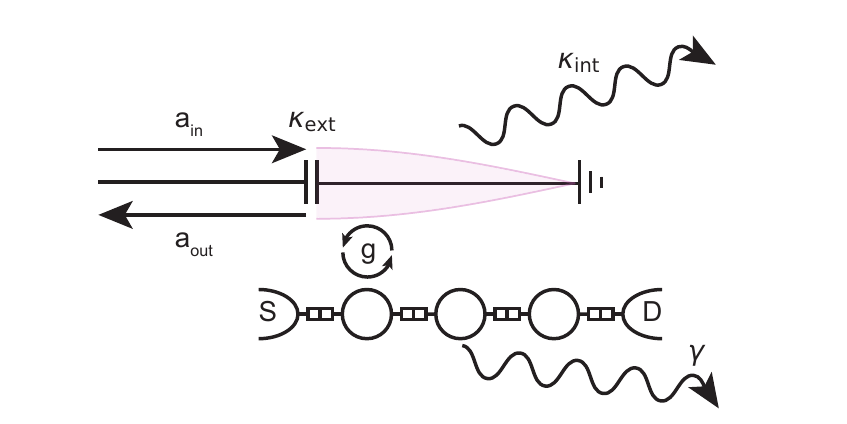}
    \caption{Schematic model of the TQD coupled to a resonator being fed by a input line. The input field $a_{\mathrm{in}}$ of the feed line at frequency $\nu_{\mathrm{p}}$ couples at rate $\kappa_{\mathrm{ext}}$ to the $\lambda/4$ resonator. The internal losses of the resonator are described by $\kappa_{\mathrm{int}}$. The resonator couples to the TQD with strength $g$. We summarize the losses of the TQD qubit with the rate $\gamma$
    \label{fig:FigS2}}
\end{figure}

In this section we discuss the details of the Input-Output model used to simulate the reflected resonator signal. We closely follow the approach presented in \cite{Burkard2016}. A schematic of the TQD coupled to a cavity including the relevant loss channels is provided in Fig.~\ref{fig:FigS2}. We consider the input field  $a_{\mathrm{in}}$ of the transmission line at frequency $\nu_{\mathrm{p}}$ coupling at rate $\kappa_{\mathrm{ext}}$ to the $\lambda/4$ resonator. We collect all internal losses of the resonator in $\kappa_{\mathrm{int}}$. The resonator and the TQD couple with strength $G$, see Sec. \ref{sec:Theory} for derivation. For our qubit this strength depends on the detuning parameter $\delta$. We summarize the losses of the TQD qubit in $\gamma$, neglecting quantum noise. We start with the system Hamiltonian $H_{\mathrm{sys}}= H_{\mathrm{C}}+H_{\mathrm{TQD}}+H_{\mathrm{int}}$ consisting of the cavity Hamiltonian $H_{\mathrm{C}}$, the TQD Hamiltonian $H_{\mathrm{TQD}}$ and the interaction Hamiltonian $H_{\mathrm{int}}$. In a second step we solve the equations of motion for the cavity and qubit operators and derive the Input-Output model. 

The triple quantum dot Hamiltonian is given by: 

\begin{equation}
    H_{\mathrm{TQD}} = \begin{pmatrix}
            \delta/2 & t_{\mathrm{L}} & 0 \\
            t_{\mathrm{L}}^\star & E_{\mathrm{M}} & t_{\mathrm{R}} \\ 
            0 & t_{\mathrm{R}}^\star & -\delta/2
        \end{pmatrix}, \label{eq:H0}
\end{equation}
The inter dot tunnel couplings of the left-middle and right-middle quantum dots are given by $t_{\mathrm{L}}$ and $t_{\mathrm{R}}$ respectively. The qubit detuning parameters $\delta$ and $E_{\mathrm{M}}$ are defined as in the main text and indicated in Fig.~\ref{fig:Fig1}(a). The resonator at resonance frequency $\omega_{\mathrm{C}}$ is described by the Hamiltonian 
\begin{equation}
    H_{\mathrm{C}} = \omega_{\mathrm{C}} a^{\dagger}a,
\end{equation}
where $a$ is the photon annihilation operator. The coupling between the two quantum systems is described by the interaction Hamiltonian $H_{\mathrm{int}}$

\begin{equation}
    H_{\mathrm{int}} = G\left(a+a^{\dagger}\right),
\end{equation}
where the coupling matrix $G$ is defined as in the main text. For further calculations it is convenient to work in the eigenbasis of $H_{\mathrm{TQD}}$. Like in the main text, let $S$ be the unitary operator that diagonalizes $H_{\mathrm{TQD}}$. With this, the qubit Hamiltonian reads 
\begin{equation}
    \tilde{H}_{\mathrm{TQD}} = SH_{\mathrm{TQD}}S^{\dagger} = \sum_{\mathrm{n}=0}^{\mathrm{n}=2} E_{\mathrm{n}}\sigma_{\mathrm{nn}},
\end{equation}
where $E_{\mathrm{i}}$ are the ordered eigenvalues of $H_{\mathrm{TQD}}$. The operator $\sigma_{\mathrm{nn}}$ is defined by $\sigma_{\mathrm{nn}} = \ket{n}\bra{n}$, where $\ket{n}$ is the eigenstate at energy $E_{\mathrm{n}}$. Under the transformation $S$ the coupling matrix transforms to: 

\begin{equation}
    \tilde{G} = SGS^{\dagger} = \sum_{\mathrm{m,n} =0}d_{\mathrm{m,n}} \sigma_{\mathrm{mn}},
\end{equation}
where $d_{\mathrm{nm}} = d_{\mathrm{mn}}^*$ are the transition matrix elements between the different eigenstates. In a next step we transform into the rotating frame of the probe frequency $\omega_{\mathrm{p}} = 2\pi \nu_{\mathrm{p}}$. The unitary transformation is given by:

\begin{equation}
    U_{\mathrm{R}}(t) = \exp\left[-it\left(\omega_{\mathrm{p}}a^{\dagger}a+\sum_{\mathrm{n} = 0}^{2} \mathrm{n} \omega_{\mathrm{p}} \sigma_{\mathrm{nn}}\right)\right].
\end{equation}
The total system Hamiltonian $\tilde{H}_{\mathrm{sys}}$ transforms as: 
\begin{equation}
    \bar{H}_{\mathrm{sys}} = U_{\mathrm{R}} \tilde{H}_{\mathrm{sys}}U_{\mathrm{R}}^{\dagger} + i \dot{U}_{\mathrm{R}}U_{\mathrm{R}}^{\dagger}
\end{equation}
Applying this transformation we find:
\begin{align}
    \bar{H}_{\mathrm{TQD}} =&\sum_{\mathrm{n}=0}^{2}\left(E_{\mathrm{n}}-\mathrm{n}\omega_{\mathrm{p}}\right)\sigma_{\mathrm{nn}}\\
    \bar{H}_{\mathrm{C}} =& \Delta_0 a^{\dagger}a\\
    \bar{H}_{\mathrm{int}} \simeq& \left(a\sum_{\mathrm{n}=0}^{2}d_{\mathrm{n}+1,\mathrm{n}}\sigma_{\mathrm{n}+1,\mathrm{n}} + \mathrm{H.c.}\right),
\end{align}
where $\Delta_0 = \omega_{\mathrm{c}}-\omega_{\mathrm{p}}$ is the detuning of the cavity frequency from the probe frequency $\omega_{p}$. We collect the dissipative losses of the system in the term  $H_{\mathrm{diss}}$. It takes the internal losses of the cavity $\kappa_{\mathrm{int}}$ and the qubit $\gamma$ to the environment into account. In the following we neglect quantum noise within the TQD. Given the Hamiltonian of the TQD system coupled to a resonator we calculate the cavity response using Input-Output theory. The equations of motion for $a$ and $\sigma_{\mathrm{n},\mathrm{n}+1}$ read as:
\begin{align}
    \dot{a} =& i\left[\bar{H}_{\mathrm{sys}}+\bar{H}_{\mathrm{diss}},a\right]\\
    \dot{\sigma}_{\mathrm{n},\mathrm{n}+1} =& i\left[\bar{H}_{\mathrm{sys}}+\bar{H}_{\mathrm{diss}},\sigma_{\mathrm{n},\mathrm{n}+1}\right]. 
\end{align}
Calculating the commutators from above we find:
\begin{align}
    \dot{a} = & -i\Delta_{\mathrm{rq}} a- i\sum_{\mathrm{n}=0}^{2} d_{\mathrm{n,n}+1}\sigma_{\mathrm{n,n}+1}\\
    &+\sqrt{\kappa_{\mathrm{ext}}}a_{\mathrm{in}} -\frac{\kappa_{\mathrm{int}}+\kappa_{\mathrm{ext}}}{2}a \label{eq:dota}\\
    \dot{\sigma}_{\mathrm{n},\mathrm{n}+1} =& -i\left(E_{\mathrm{n}+1}-E_{\mathrm{n}} - \omega_{\mathrm{p}}\right) \sigma_{\mathrm{n},\mathrm{n}+1}\\ 
    & -i d_{\mathrm{n}+1,\mathrm{n}}\left(p_{\mathrm{n}}-p_{\mathrm{n}+1}\right)a 
    - \frac{\gamma}{2}\sigma_{\mathrm{n,n}+1}, \label{eq:dotsigma}
\end{align}
where $p_{\mathrm{n}}$ is the occupation probability of state $\mathrm{n}$, coming from evaluating terms of the form $\left[\sigma_{\mathrm{n}+1,\mathrm{n}},\sigma_{\mathrm{n,n}+1}\right]$. In thermal equilibrium the occupation probability is described by Boltzmann statistics
\begin{equation}
    p_{\mathrm{n}} = \frac{\exp\left(-E_{\mathrm{n}}/k_{\mathrm{B}}T\right)}{\sum_{\mathrm{n}} \exp\left(-E_{\mathrm{n}}/k_{\mathrm{B}}T\right)}.
\end{equation}
Solving Eq.~(\ref{eq:dota}) in the stationary limit we find an expression for $a/\sigma_{\mathrm{n,n}+1} = \chi_{\mathrm{n,n}+1}$,
\begin{equation}
    \sigma_{\mathrm{n,n}+1} = \frac{-d_{\mathrm{n,n}+1}\left(p_{\mathrm{n}}-p_{\mathrm{n}+1}\right)}{E_{\mathrm{n}+1}-E_{\mathrm{n}}-\omega_{\mathrm{p}} - i \gamma/2} =\chi_{\mathrm{n,n}+1} a
\end{equation}
Solving Eq.~(\ref{eq:dotsigma}) and using the above expression we find an equation for $a/a_{\mathrm{in}}$ \cite{Gardiner1984}:
\begin{equation}
\frac{a}{a_{\mathrm{in}}}=\frac{i \sqrt{\kappa_{\mathrm{ext}}}}{\omega_{\mathrm{c}}-\omega_{\mathrm{p}}- i \left(\frac{\kappa_{\mathrm{int}}+\kappa_{\mathrm{ext}}}{2}\right)+  \sum_{\mathrm{n}=0}^{2} d_{\mathrm{n,n}+1} \chi_{\mathrm{n,n}+1}}
\end{equation}
Taking into account that we use a $\lambda/4$ cavity we find the relation 
\begin{equation}
    a_{\mathrm{out}} = \sqrt{\kappa_{\mathrm{ext}}} a - a_{\mathrm{in}}
\end{equation}
between the output field and the input field operators.
Using this expression we get the coefficient
\begin{equation}
    A = \frac{a_{\mathrm{out}}}{a_{\mathrm{in}}}=\frac{\omega_{\mathrm{p}}-\omega{\mathrm{c}}+i\frac{\kappa_{\mathrm{int}}-\kappa_{\mathrm{ext}}}{2} +  \sum_{\mathrm{n}=0}^{2} d_{\mathrm{n,n}+1} \chi_{\mathrm{n,n}+1} }{\omega_{\mathrm{c}}-\omega_{\mathrm{p}}- i
    \frac{\kappa_{\mathrm{int}}+\kappa_{\mathrm{ext}}}{2}+ \sum_{\mathrm{n}=0}^{2} d_{\mathrm{n,n}+1} \chi_{\mathrm{n,n}+1}}
\end{equation}
This coefficient is related to the measured reflectivity by $\left|S_{11}\right|= |A|^2$.

\subsection{Vacuum Rabi Splitting}

\begin{figure}
    \centering
    \includegraphics[width=\columnwidth]{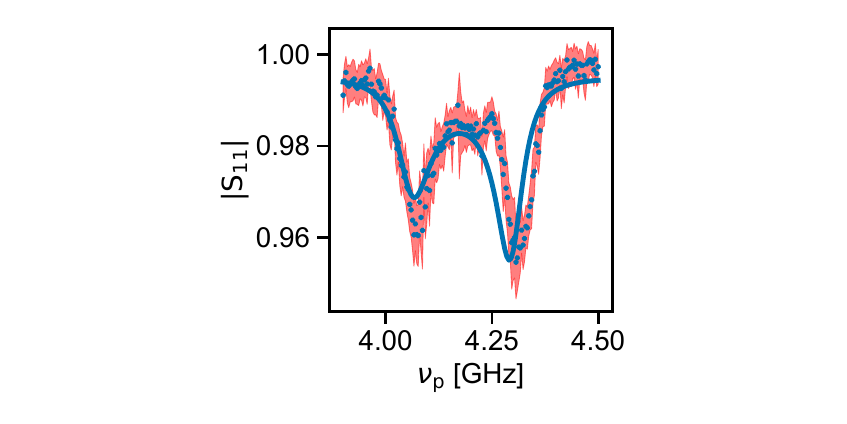}
    \caption{Vacuum Rabi cut at $\delta = 0$ from the measurement presented in Fig.~\ref{fig:Fig2}(d). We measured the linecut 40 times consecutively, blue points correspond to the averaged data points, the red area shows the standard deviation of the averages. The solid line is the corresponding linecut taken from Fig.~\ref{fig:Fig2}(e).
    \label{fig:FigS4}}
\end{figure}

Additionally to the measurement presented in Fig.~\ref{fig:Fig2}(d), we present the vacuum Rabi cut at $\delta = 0$ in Fig.~\ref{fig:FigS4}. For this linecut we perform 40 repetitions. The dots represent the average of the 40 linecuts. The red region represents the standard deviation of the average. The solid blue line is taken from the Input-Output model calculations presented in Fig.~\ref{fig:Fig2}(e) directly, not doing any separate fitting. One can clearly resolve the two peaks in the vacuum Rabi splitting and therefore the strong coupling.

\subsection{Photon Number calibration}

In the following section we discuss how we calibrate the photon number in the resonator. The strong coupling of a qubit to a cavity radiation file leads to a dressed qubit state whose energy depends on the occupation number $n$ of the resonator \cite{Schuster2005}. The dressed qubit frequency $\tilde{\nu}_{\mathrm{q}}$ is given as: 
\begin{equation}
    \tilde{\nu}_{\mathrm{q}} = \nu_{q}  + 2 \frac{n \left(g/2\pi\right)^2}{\Delta_{\mathrm{qr}}} + \frac{ \left(g/2\pi\right)^2}{\Delta_{\mathrm{qr}}},
\end{equation}
where $\Delta_{\mathrm{qr}}= \nu_{\mathrm{q}}-\nu_{\mathrm{r}}$ is the qubit cavity detuning. In this experiment we use the DQD charge qubit formed between the left-middle DQDs to calibrate the photon number $n$. Changing the flux $\Phi$ we tune the resonance frequency of the resonator to $\nu_{\mathrm{r}} = \SI{4.2}{\giga\hertz}$. The qubit frequency is given by $2t \approx \SI{5}{\giga\hertz}$. Using two tone spectroscopy we measure the qubit frequency $\tilde{\nu_{\mathrm{q}}}$ while increasing the applied probe tone power to the resonator. We find a linear change of $\tilde{\nu}_{\mathrm{q}}$. We determine the slope $a$ by linear fit. From this we find:
\begin{equation}
    n/P = \frac{a\Delta_{\mathrm{qr}}}{2\left(g/2\pi\right)^2}.
\end{equation}
In the experiment we apply a resonator power corresponding to $n \approx 0.05$. 

\begin{figure}
    \centering
    \includegraphics[width=\columnwidth]{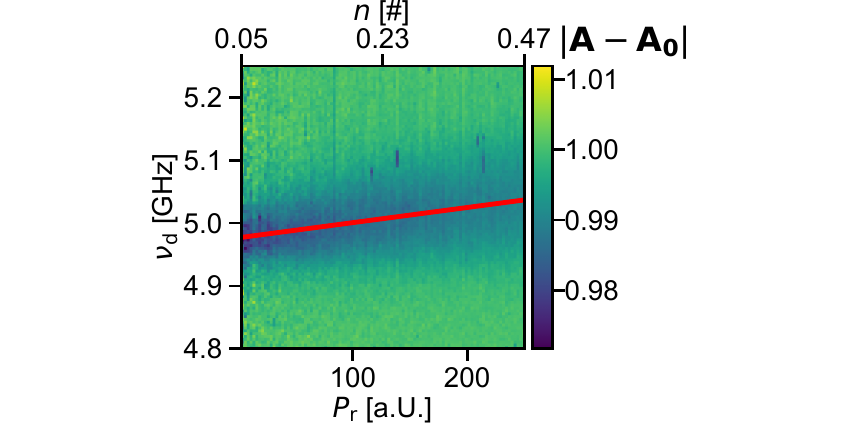}
    \caption{AC-Stark shift measurement performed on the left-middle DQD charge qubit. We measure the dressed qubit frequency $\tilde{\nu}_{\mathrm{q}}$ as a function of applied resonator probe tone power. We observe a linear shift of $\tilde{\nu}_{\mathrm{q}}$. From this shift we calculate the average photon number occupation in the resonator $n$ as indicated in the second x-Axis.
    \label{fig:FigS5}}
\end{figure}

\subsection{Sweet spot characterization}
In previous studies with charge-quadrupole qubits the information is encoded in the ground and second excited states, allowing to work in a sweet spot for both dipolar and quadrupolar detunings \citep{Friesen2017, Koski2019}. In the $CQ_3$ qubit, however, the qubit is encoded in the two lowest states, such that the sensitivity to quadrupolar charge noise is sacrificed in favor of an increased insensitivity to dipolar charge noise. This alternative encoding, gives rise to a different sweet spot landscape which is characterized in this section.

\begin{figure}
    \centering
    \includegraphics[width=\columnwidth]{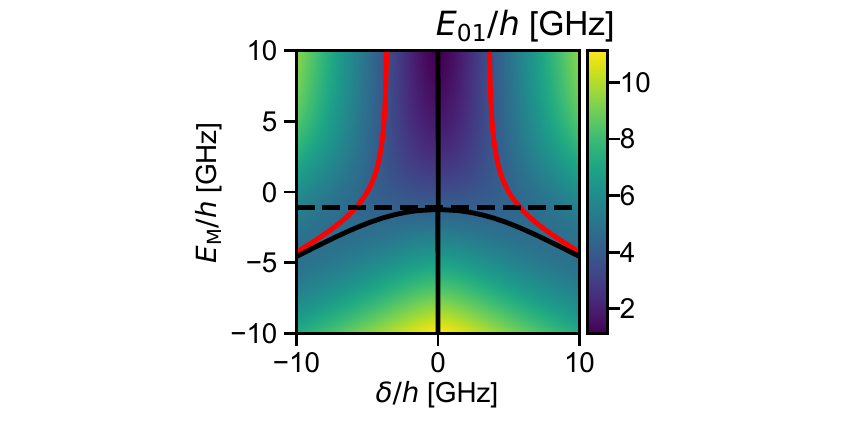}
    \caption{Plot of the qubit energy $E_{01}/h$ as a function of $\delta$ and $E_{\mathrm{M}}$ for $|t| = |t_{\mathrm{L}}| = |t_{\mathrm{R}}| = \SI{2.5}{\giga\hertz}$. Red lines indicate the position of sweet spots in $E_{\mathrm{M}}$ and the black solid line indicates sweet spots in $\delta$. The black dashed line shows the $E_{\mathrm{M}}^{Opt} \approx -0.493t$.
    \label{fig:FigS7}}
\end{figure}

In Fig.~\ref{fig:FigS7}, the qubit energy is plotted as a function of the $\delta$ and $E_{\mathrm{M}}$ for $|t_{\mathrm{L}}|=|t_{\mathrm{R}}| = \SI{2.5}{\giga\hertz}$. The black and red contours show the curves in which there is a dipolar and quadrupolar sweet spot, respectively. It can be seen that, for this encoding, simultaneous sweet spots of dipolar and quadrupolar detunings cannot exist. The optimal working point as referred to in the main text, occurs when the two black curves cross, at $\delta=0$ and $E_{\mathrm{M}}\approx -0.493t$. We note that, the condition $|t|=|t_{\mathrm{L}}|=|t_{\mathrm{R}}|$ is necessary for the second-order sweet spot to occur.

\subsection{Noise Model}

The qubit decoherence rate $\gamma_2$ is obtained by its relation with the HWHM $\partial \nu_{\mathrm{q}}$ obtained for a given power $P$: 
\begin{equation}
    \partial\nu_{\mathrm{q}}=\sqrt{(\gamma_2/2\pi)^2+\beta P}
\end{equation}
where $\beta$ is a constant that is calibrated as previously explained, see discussion of Fig.~\ref{fig:Fig4} in the main text. We model the qubit decoherence rate by simulating Ramsey free induction decay in the $CQ_3$ qubit. We expect the qubit decoherence rate to be dominated by charge fluctuations in the different detuning parameters, and Overhauser magnetic fluctuations.

\textbf{Magnetic noise.}
While magnetic noise is not the dominant decoherence mechanism, its impact on the qubit is non-negligible. The magnetic noise fluctuations in GaAs are slow compared to the qubit dynamics, hence, allowing to assume a quasistatic Gaussian noise distribution. Following Ref. \onlinecite{Fei2015} we name the difference in Overhauser fields between left-middle dots and middle-right dots as $B_{\mathrm{L}}$ and $B_{\mathrm{R}}$, respectively. We note that, since this is a charge-type of qubit, a global shift in the magnetic field has a negligible influence on the qubit coherence, allowing to characterize the magnetic noise fluctuations in three dots with two parameters. 

\begin{figure}
    \centering
    \includegraphics[width=\columnwidth]{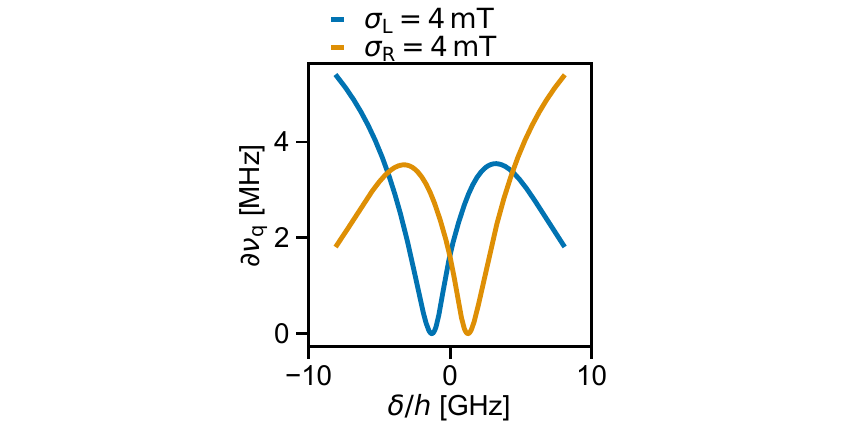}
    \caption{Simulated influence of magnetic noise arising from different Overhauser fields in different dots. We assume Gaussian distributed magnetic noise with a standard deviation of $\sigma = \SI{4}{\milli\tesla}$ on either the left-middle or middle-right DQD. Note the different axis scales compared to Fig.~\ref{fig:Fig4}(c).
    \label{fig:FigS6}}
\end{figure}

The Hamiltonian due to Overhauser fields is
\begin{equation}
    H=\frac{\mathrm{g}\mu_B}{6}\begin{pmatrix}
        2B_{\mathrm{L}}+B_{\mathrm{R}} & 0 & 0 \\
        0 & B_{\mathrm{L}}-B_{\mathrm{R}} & 0 \\
        0 & 0 & -B_{\mathrm{L}}-2B_{\mathrm{R}}
    \end{pmatrix}.
\end{equation}
Assuming a Gaussian distribution of local magnetic fields with standard deviations $\sigma_{\mathrm{L}}$ and $\sigma_{\mathrm{R}}$, the decoherence rate is \citep{Landig2019}
\begin{equation}
\frac{\gamma_2}{2\pi}=\mathrm{g}\mu_B\sqrt{\frac{h_{\mathrm{L}}^2\sigma_{\mathrm{L}}^2+h_{\mathrm{R}}^2\sigma_{\mathrm{R}}^2}{2}},
\end{equation}
where $h_{\mathrm{L,R}}=d\omega_{01}/dB_{\mathrm{L,R}}$, being $\omega_{01}$ the qubit frequency.  The result of using the previous formula for $\sigma_{\mathrm{L}}=\SI{4}{\milli\tesla}$, $\sigma_{\mathrm{R}}=0$, and vice versa, is shown in Fig.~\ref{fig:FigS6}. In GaAs the fluctuations of the nuclear magnetic field are expected to be around \SIrange{2}{5}{\milli\tesla}. The asymmetry of the results in Fig.~\ref{fig:Fig4}(b) is not caused by these fluctuations. Moreover, due to the small value of the decoherence rate due to this mechanism compared to the observed values, we assume a typical value of $\sigma_{\mathrm{L}}=\sigma_{\mathrm{R}}=\SI{4}{\milli\tesla}$ in the following. Realistic deviations from this value would be negligible compared to the numbers in Fig.~\ref{fig:Fig4}(b).

\textbf{Charge noise.}
We assume the overall charge fluctuations follow a $1/f$ spectral distribution $S_i(\omega)=2\pi A_i/\omega$, where $i$ indicates parameter $i=\delta,E_{\mathrm{M}}$, and $A_i$ is its corresponding noise amplitude. These fluctuations induce variations in the chemical potentials of the different dots, such that $\delta\rightarrow \delta+\Delta\delta(t)$, and $E_{\mathrm{M}}\rightarrow E_{\mathrm{M}}+\Delta E_{\mathrm{M}}(t)$. For fitting the observed decoherence rate, we consider three parameters: the dipolar detuning noise amplitude $A_\delta$, the quadrupolar detuning noise amplitude $A_{E_{\mathrm{M}}}$ and the correlation between each dipolar and quadrupolar fluctuations $\rho$. This coefficient allows the dipolar and quadrupolar fluctuations to be correlated. To account for the asymmetry observed in Fig.~\ref{fig:Fig4}(b) we assume that the strength of the qubit drive tone experienced by the qubit decays with coupling to the drive gate.

For given values of the three noise parameters in a certain qubit configuration $(\delta,E_{\mathrm{M}})$ the procedure to obtain the value of $\gamma_2$ goes as follows:
\begin{itemize}
    \item The qubit is initialized in a coherent superposition $\ket{\psi}=1/\sqrt{2}(\ket{0}+\ket{1})$
    \item Noise fluctuations $\Delta\delta(t)$ and $\Delta E_{\mathrm{M}}(t)$ are generated following the method described in Refs. \onlinecite{Yang2019, Koski2019}.
    \item The qubit is left to evolve under the noise fluctuations for a time $\tau=100$ ns. 
    \item The evolution of the qubit coherence $\rho_{01}(t)$ is saved.
    \item The previous steps are repeated 5000 times. 
    \item The multiple resulting evolutions of the qubit coherence are averaged. 
    \item The value of $|\rho_{01}(t)|$ follows a decay law $|\rho_{01}(t)|=\exp(-\gamma_2 t/2\pi)^\beta/2$. Fitting to such function provides the decay rate $\gamma_2$ and the exponent $\beta$.
\end{itemize}
The result of applying this procedure to the simple cases $A_\delta=\SI{1}{\micro\eV}$, $A_{E_{\mathrm{M}}}=0$, and vice versa, are shown in Fig.~\ref{fig:Fig4}(c).
To fit the result in Fig.~\ref{fig:Fig4}(b), this procedure is repeated over a grid in $\Delta\delta$, $A_\delta$, $A_{E_{\mathrm{M}}}$, and $\rho$ for $\Delta E_{\mathrm{M}}=E_{\mathrm{M}}^{\text{Opt}}$. This grid is then interpolated into a function to which we add the decoherence rate from magnetic fluctuations. The interpolated function is then used to fit Fig.~\ref{fig:Fig4}(b). The result of the fit gives $A_{\delta} = \SI{1.949\pm 0.098}{\micro\eV}, A_{E_{\mathrm{M}}} = \SI{0.935\pm0.026}{\micro\eV}, \mathrm{corr} = \SI{-0.084\pm 0.177}{},$ with a shift in detuning $\delta' = \SI{0.69\pm0.171}{\micro\eV}$. This implies that the dipolar detuning noise is dominating over the quadrupolar noise, in a similar ratio to the one observed in a previous work \citep{Koski2019}.

\bibliographystyle{apsrev4-2}
\bibliography{references}
\end{document}